\documentclass[trackchanges]{aastex701}
\usepackage{amsmath}
\begin{document}

\title{A Self-Consistent Model of the Ultra High-Energy Gamma-Ray Emission of Pulsar Wind Nebulae: Insights from LHAASO and ATNF Catalogs}

\author[0009-0008-1503-7056]{Samy Kaci}
\email[show]{samykaci@sjtu.edu.cn}
\correspondingauthor{Samy Kaci}
\affiliation{Tsung-Dao Lee Institute, Shanghai Jiao Tong University, Shanghai 201210, P. R. China}
\affiliation{School of Physics and Astronomy, Shanghai Jiao Tong University, Shanghai 200240, P. R. China}

\author[0000-0001-9745-5738]{Gwenael Giacinti}
\email[show]{gwenael.giacinti@sjtu.edu.cn}
\correspondingauthor{Gwenael Giacinti}
\affiliation{Tsung-Dao Lee Institute, Shanghai Jiao Tong University, Shanghai 201210, P. R. China}
\affiliation{School of Physics and Astronomy, Shanghai Jiao Tong University, Shanghai 200240, P. R. China}
\affiliation{Key Laboratory for Particle Physics, Astrophysics and Cosmology (Ministry of Education) \& Shanghai Key Laboratory for Particle Physics and Cosmology, 800 Dongchuan Road, Shanghai, 200240, P. R. China}

\author[0000-0001-6417-1560]{Dmitri Semikoz}
\affiliation{APC, Université Paris Cité, CNRS/IN2P3, CEA/IRFU, Observatoire de Paris, 119 75205 Paris, France}
\email{dmitri.semikoz@apc.univ-paris7.fr}

\begin{abstract}
Pulsar wind nebulae (PWNe) are the dominant Ultra-high-energy (UHE) gamma-ray sources in the LHAASO catalog suggesting that they are the dominant leptonic PeVatrons in our Galaxy. Despite this, still very little is known about their UHE gamma-ray emission, their number in the Galaxy, or their contribution to the gamma-ray emission of our Galaxy. In this work, we propose a self-consistent data-driven model of the UHE gamma-ray emission of PWNe based on the ATNF and LHAASO catalogs. More specifically, we build a model of the UHE gamma-ray emission of PWNe that preserves the statistical relationships in the ATNF catalog and reproduces the number of PWNe detected in the LHAASO catalog. To cope with the limited data available in the LHAASO catalog when performing fits on gamma-ray data, we introduce the concept of censored regression that allows to also use the information provided by unresolved sources. Using our model, we find that reproducing the number of PWNe detected by LHAASO requires either fractions of misaligned pulsars smaller ($\lesssim60\%$) than usually found in the literature, or that some of the associations of PWNe to ATNF pulsars made by LHAASO may not be true. In both cases, we find that in order to reach self-consistency between radio and gamma-ray data, it is necessary that the majority of the unidentified sources in the LHAASO catalog are PWNe associated to an unseen pulsar. Moreover, using our model we also find that the contribution of unresolved PWNe to the total (diffuse) gamma-ray background measured by LHAASO in the $1\textendash1000\,\rm{TeV}$ range is always smaller than $\lesssim10\%$ ($\lesssim30\%$). We conclude that PWNe mostly contribute to the source component of the UHE gamma-ray sky, while having almost no imprint on its diffuse component.

\end{abstract}


\section{Introduction} \label{sec:intro}
In recent years the new generation of ground-based gamma-ray observatories, such as AS$\gamma$, HAWC or LHAASO, have opened the ultra-high-energy (UHE) observational window offering precise measurements of the UHE diffuse gamma-ray background \citep{lhaaso_diffuse, lhaaso_diffuse_2} or the first LHAASO catalog of gamma-ray sources \citep{lhaaso_catalog}. While the quest of hadronic PeVatrons is still on-going, it has become clear from the current data that Pulsar Wind Nebulae (PWNe), the dominant gamma-ray sources in the LHAASO catalog, are most likely the main leptonic PeVatrons in our Galaxy, capable of achieving particle acceleration up to multi-PeV energies \citep{crab_pev}. On the other hand, despite the accumulation of observational data, the gamma-ray emission of such objects remains poorly understood and very challenging to predict. Indeed, the gamma-ray emission of PWNe strongly depends on the specifics of the particle acceleration and gamma-ray production processes taking place locally. More specifically, the UHE gamma-ray emission of PWNe, which is primarly due to inverse Compton (IC) scattering of relativistic electrons and positrons on photons, directly depends on the local (and poorly known) photon fields as well as the local density of leptons and their energy spectrum. In turn, these heavily depend on the unknown acceleration mechanism, and on the multiplicity factor of the parent pulsar wind. Because of all these unknowns, it becomes nearly impossible to build a satisfying model of the gamma-ray emission of PWNe from first principles. On the other hand, the derivation of such a model is of utmost importance for the modeling of the UHE Galactic gamma-ray emission. In particular, a realistic model of the UHE gamma-ray emission of PWNe is crucial to constrain the number of leptonic PeVatrons in our Galaxy and their contribution to the source and to the diffuse components of the Galactic gamma-ray flux \citep{me2, neutrinos_tevhalo, tevhalos_let}.

In this work, we propose a new data-driven procedure that aims to provide a phenomenological model of the UHE gamma-ray emission of PWNe by simultaneously accounting for the various observations in gamma-rays, such as the number of detectable sources, without violating radio observations in the ATNF\footnote{\url{https://www.atnf.csiro.au/research/pulsar/psrcat/}} catalog \citep{atnf_1}. To do that, we thoroughly revisit the LHAASO associations of gamma-ray sources to known pulsars \citep{lhaaso_catalog} by adapting to PWNe the procedure introduced by \cite{Mattox_1997} to associate EGRET gamma-ray sources with extragalactic radio sources. This step is necessary to ensure the reliability of our results. Indeed, because of the very limited statistics, the best-fit values are sensitive to the fitted data and, therefore, to unreliable associations of LHAASO sources with ATNF pulsars. Moreover, in order to better cope with the very limited statistic in the UHE band, we introduce the concept of censored linear regression, also known as Tobit regression model, which enables the use of the information carried by the detected sources, for which a gamma-ray spectrum is available, and also the information carried by the undetected sources, through upper limits on their flux.

This paper is organized as follows: In Section \ref{sec:methods}, we present our model to generate synthetic pulsar populations and introduce the concept of censored regression that we use to generate the gamma-ray spectra of the PWNe associated to the synthetic pulsars. In section \ref{sec:off_beamed_sec}, we present our adaptation of the procedure of \cite{Mattox_1997}, we use it to reexamine the associations made by LHAASO and finally infer the parameters of our model of the UHE gamma-ray emission of PWNe. In Section \ref{sec:background}, we use our model to evaluate the contribution of unresolved PWNe to the total and to the diffuse gamma-ray fluxes measured by LHAASO. Finally, we present our discussion and conclusions in Section \ref{sec:conclusion}. For clarity, in what follows we present and describe all our procedures only for the KM2A detector of LHAASO, the extension of our procedures to to the WCDA detector being straightforward.

\section{Methods} \label{sec:methods}
In this paper, we aim to provide a self-consistent model of the UHE gamma-ray emission of PWNe in our Galaxy. The construction of such a model requires to: First, define and model the parent pulsar population and their physical parameters, and second, build a model for the UHE gamma-ray emission of the PWNe associated to the parent pulsar population. In this section, we present our procedure to generate the population of pulsars and the UHE gamma-ray emission of their PWNe.

\subsection{The Parent pulsar population} \label{subsec:generation}

To define our model of the pulsar population, it is sufficient to know the spatial distribution of pulsars in the Galaxy, assumed to be axi-symmetric, their age distribution and the relationship between the spindown power of pulsars and their age. For that, we use the ATNF catalog and mostly follow, with some improvements, the approach presented in \cite{me2} that we summarize here.

\begin{figure}
    \centering
    \includegraphics[width=\linewidth]{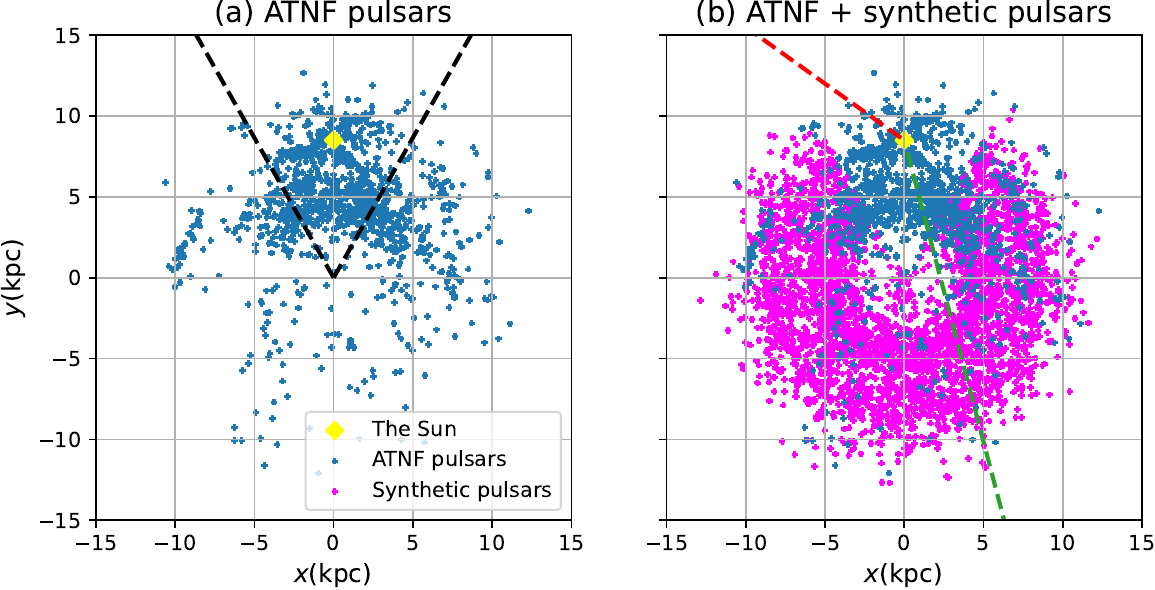}
    \caption{Panel (a) shows the locations of radio-loud pulsars younger than $10\,\rm{Myr}$ with altitudes $|z|\leq1.7\,\rm{kpc}$ in the ATNF catalog. The pulsars between the two dashed lines in black that form an angle of $60^{\circ}$ at the Galactic center have been used to extract the local radial and age distributions of pulsars, assumed to be representative of the entire Galaxy. Panel (b) shows the locations radio-loud pulsars in the ATNF catalog together with the synthetic pulsars generated following our approach. The field of view of LHAASO goes counter-clockwise from the green dashed line to the red dashed line. In both panels the Sun is represented by a yellow diamond.}
    \label{fig:distributions}
\end{figure}

In order to obtain the radial and age distributions of pulsars, we first select from the ATNF catalog only pulsars that have a measured distance to the Sun and with altitudes $|z|\leq1.7\,\rm{kpc}$. This cut in altitude corresponds to the region of interest (ROI) in \cite{lhaaso_diffuse, lhaaso_diffuse_2}. In addition, we apply an age cut and only keep pulsars with ages smaller than $10\,\rm{Myr}$. Older pulsars can be safely discarded in virtue of the Hillas criterion \citep{e_max} as explained in the next subsection. Moreover, we make the conservative assumption that all radio-quiet pulsars are misaligned pulsars\footnote{It should be noted that the ATNF catalog includes all pulsars that have measurements for both of their rotation period $P$ and its derivative $\dot{P}$. As a result, it includes misaligned pulsars which have not been detected in radio, but have been detected in gamma-rays or in other wavelengths, such as the Fermi pulsars.} and exclude them, for now, from our sample. Finally, we consider that the ATNF catalog is sufficiently complete (excluding misaligned pulsars) inside the sector delimited by the two black dashed lines forming an angle of $60^{\circ}$ in the Galactic center and encompassing the Sun, as represented in panel (a) of figure \ref{fig:distributions}. This figure represents a view of our Galaxy from above with the ATNF pulsars being shown in cyan and the Sun in yellow.

Constraining ourselves to the sector of the panel (a) of figure \ref{fig:distributions} that we divide into $13$ radial bins, and applying the selection criteria previously enunciated we extract the number of pulsars in each bin. This procedure leads to a total number of $704$ pulsars in this portion of the Galaxy. For the age distribution we divide the pulsars into $24$ age bins and apply the same criteria as for the radial distribution.

In order to include misaligned pulsars we follow the modeling of \cite{off_beamed} which provides an expression (see equation (16) therein) relating the age of pulsars to the beaming fraction (assumed to be age-dependent).  For each age bin $\tau_i$ among the $24$ age bins, we compute the fraction of aligned pulsars $\epsilon_i$. After that, we adjust the number of pulsars in each age bin by dividing by its beaming fraction $\epsilon_i$ and then rescale the radial distribution and the total number of pulsars accordingly. At this point, we have obtained the final radial and age distribution in the sector represented in panel (a) of figure \ref{fig:distributions}. This sector represents one sixth of the Galaxy. 

In order to generate the full synthetic populations of pulsars, we divide the Galaxy into six sectors similar to the one represented in panel (a) of figure \ref{fig:distributions}. After that, we fill each sector with ATNF pulsars and supplement with synthetic pulsars following the radial and age distributions previously derived until reaching the total number of pulsars expected in each sector. To be more realistic, we allow the total number of pulsars expected in each sector, that is $704/\epsilon$ to fluctuate stochastically by some number of pulsars that is drawn randomly from a Gaussian distribution centered around $0$ with a standard deviation of $\sqrt{704/\varepsilon}$. The resulting population is shown in panel (b) of figure \ref{fig:distributions} for $\epsilon=1$ together with the field of view (FOV) of LHAASO that goes counter-clockwise from the green to red dashed line.

Finally, to generate the spindown power of synthetic pulsars, we make a linear fit of the spindown power of pulsars as a function of their ages in the ATNF catalog and extract the following expression:
\begin{equation}\label{edot}
    \log\left(\frac{\dot{E}}{\rm{erg}\,\rm{s}^{-1}}\right) = -1.529\log\left(\frac{\tau}{\rm{yr}}\right)+42.560
\end{equation}
where $\dot{E}$ is the spindown power of the pulsar and $\tau$ its age. The scattering in the data is accounted for by adding a Gaussian noise with a standard deviation $\sigma=0.699$ which represent the extent of the $68\%$ containment region of the data around the fitted line.

In summary, in our modeling we first extract the spatial and age distributions of pulsars in the vicinity of the Sun from the ATNF catalog. We then adjust these distributions to account for misaligned pulsars following \cite{off_beamed}. After that, we generate our source population by filling the Galaxy with ATNF pulsars supplemented by synthetic ones, until the expected number of pulsars in the Galaxy is reached, while allowing for stochastic fluctuations in their number from one region to another. Finally, we assign a spindown power to each synthetic pulsar using equation (\ref{edot}) which has been derived through a linear regression on the ATNF data.

\subsection{The gamma-ray emission of Pulsar Wind Nebulae}
Modeling the UHE gamma-ray emission of PWNe requires to: Define a gamma-ray spectrum whose parameters are to be determined, the maximum energy of sources, and the physical extension of their gamma-ray emission. The physical extension of the gamma-ray emission does not have a significant impact on what follows \cite{me2}. Indeed, for a detector on Earth only the angular extension of a source, which depends on both the distance of the source to the Sun and its extension, is relevant. Therefore, we take a common extension of $20\,\rm{pc}$ for all sources, which matches the extension of Geminga and Monogem \citep{gg}. For the maximum energy achievable in each PWNe, it can be inferred through a Hillas-like-criterion \citep{e_max} and is given, in the most optimistic case, by
\begin{equation}\label{e_max_photon}
    E_{\gamma,\rm{max}} \approx 0.9\dot{E}_{36}^{0.65}\,\rm{PeV}
\end{equation}
where $\dot{E}_{36}$ is the spindown power of the pulsar in units of $10^{36}\,\rm{erg}\,\rm{s}^{-1}$. It should be noted that equations (\ref{edot}) and (\ref{e_max_photon}) impose a maximum energy of $\sim1\,\rm{TeV}$ on average for the gamma-ray emission of a PWN whose age is $\sim10\,\rm{Myr}$, which is at the lower end of the energy range we are interested in and justifies our choice to exclude pulsars older than $10\,\rm{Myr}$ in subsection \ref{subsec:generation}.

Finally, to determine the spectrum of sources we follow \cite{lhaaso_catalog} and assume a simple power-law gamma-ray spectrum which reads as
\begin{equation}\label{sepctrum_def}
    dN/dE \equiv N_0\left(E/E_0\right)^{-\Gamma}
\end{equation}
where $N_0$ is the reference flux at the reference energy $E_0 = 50\,\rm{TeV}$ for the KM2A detector and $\Gamma$ is the spectral index of the source.

In the context of equation (\ref{sepctrum_def}), the gamma-ray spectrum of a source is entirely defined by $N_0$ and $\Gamma$ which are to be determined. To do so, we propose the following procedure. We first fit the total flux emitted above $10\,\rm{TeV}$ by the source as a function of its spindown power. The total flux emitted above $10\,\rm{TeV}$ by a source reads as
\begin{equation}\label{total_flux}
    4\pi d^2F\left(E>10\,\rm{TeV}\right) = 4\pi d^2\int_{10\,\rm{TeV}}^{E_{\rm{max}}}\frac{dN}{dE}dE
\end{equation}
where $d$ is the distance between the source and the Sun, $dN/dE$ is defined by equation (\ref{sepctrum_def}) and $E_{\rm{max}}$ by equation (\ref{e_max_photon}). The quantity $4\pi d^2F\left(E>10\,\rm{TeV}\right)$ represents the intrinsic emission of the source assumed to be isotropic and which is a meaningful physical quantity. Once the total gamma-ray flux of a source above $10\,\rm{TeV}$ is known, which is the case with equation (\ref{total_flux}), the value of either the spectral index $\Gamma$ or the reference flux $N_0$ is sufficient to entirely define the gamma-ray emission of the source, with the determination of the value of the other parameter being straightforward. Here, we choose to randomly generate the spectral index of the synthetic source following the distribution of spectral indices of the $31$ PWNe reported by KM2A in \cite{lhaaso_catalog} and then infer the reference flux. The underlying reason behind this choice is that the distribution of spectral indices is better defined and constrained than that of the references fluxes. Choosing the reference flux first might lead to cases with an excessively hard or soft spectrum, not physically motivated and challenging to handle properly.

\begin{figure}[h!]
    \centering
    \includegraphics[width=\linewidth]{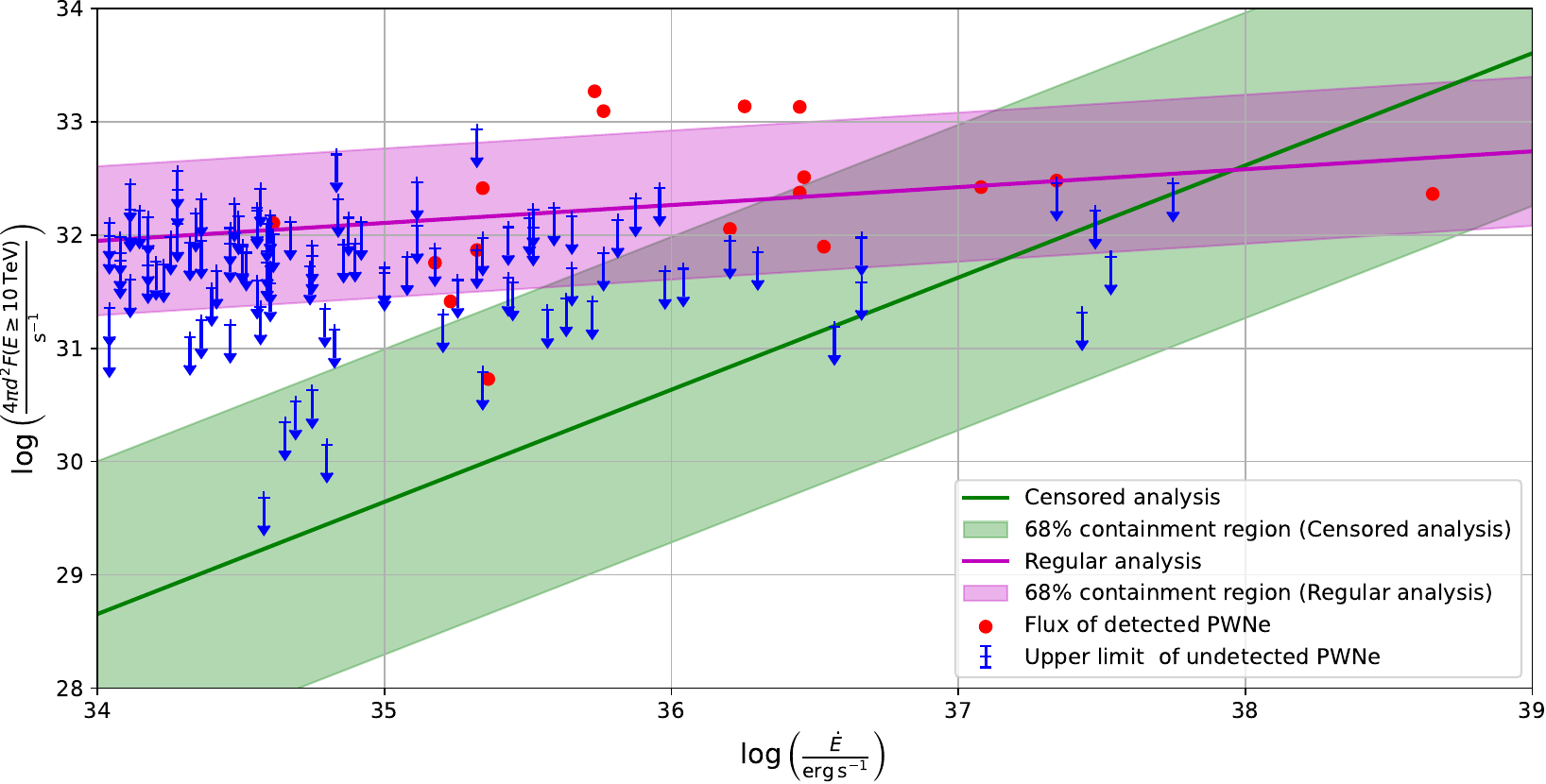}
    \caption{Result of the standard (magneta line) and censored (green line) analyses performed on the ATNF and LHAASO data. The red points represent the intrinsic emission of the PWNe detected by KM2A and the blue points represent the upper limit on the intrinsic emission of undetected PWNe. The shaded areas represent the $68\%$ containment regions resulting from the fits.}
    \label{fig:fits}
\end{figure}

The main challenge in the approach presented above lies the very limited statistics. Indeed, while several hundreds of ATNF pulsars are present in the FOV of LHAASO, only $31$ PWNe have been detected by KM2A. As a result, it becomes very difficult to fully capture the statistical relationship between the spindown power of the parent pulsars and the gamma-ray flux of the associated PWNe. In this case, any fitting attempt which relies on the small minority of detected PWNe, which are necessarily among the brightest in the Galaxy, will inevitably produce a model of the gamma-ray emission strongly biased towards these sources, failing to reproduce the gamma-ray emission of the entire population of PWNe (see figure \ref{fig:fits}).
To circumvent the limitations imposed by the very small statistics, instead of performing the fit solely on the PWNe detected by KM2A, as it is usually the case, we use a censored linear regression that uses the information from both detected and undetected (unresolved) sources. In this case the likelihood function can be written as:
\begin{equation}
    \mathcal{L}\left(\theta\right)\equiv\mathcal{L}_{\rm{detected}}\left(\theta\right)\mathcal{L}_{\rm{undetected}}\left(\theta\right)
\end{equation}
where $\mathcal{L}_{\rm{detected}}\left(\theta\right) = \prod_{i=0}^{N}P\left(y_i|x_i,\theta\right)$ and $\mathcal{L}_{\rm{undetected}}\left(\theta\right) = \prod_{j=0}^{M}P\left(y_j\leq y_{\rm{lim}}|x_j,\theta\right)$. Here $\theta = \left(\alpha, \beta,\sigma\right)$, with $\alpha$ being the slope of the linear fit, $\beta$ the y-intercept of the line and $\sigma$ the level of scattering in the data. The probability $P$ is assumed to be a Gaussian where different $x_i$ represent the spindown power of the different ATNF pulsars, $y_i$ the intrinsic emission of detected PWNe and $y_{\rm{lim}}$ the upper limit on the intrinsic emission of unresolved PWNe to ensure they remain undetected. In other words, the likelihood function for the detected sources $\mathcal{L}_{\rm{detected}}\left(\theta\right)$ is the usual one, where each factor in the product represents the probability of occurrence of each data point for a given choice of parameters $\left(\alpha, \beta,\sigma\right)$ and the likelihood function for the undetected sources $\mathcal{L}_{\rm{undetected}}\left(\theta\right)$ is written as a product where the $j$-th factor is the (cumulative) probability for the $j$-th PWN to have a flux $y_j$ below $y_{\rm{lim}}$ which is set to be the sensitivity of KM2A times $4\pi d^2$.

To ensure fairness between the two populations of detected and undetected sources, we exclude from the fit the ATNF pulsars that are misaligned and the PWNe detected by KM2A that are associated to mislaigned pulsars. We also exclude pulsars and PWNe without a measured distance, as the uncertainty on their flux/upper limit can be as large as a factor of several hundreds. The exclusion of PWNe associated to misaligned pulsars ensures fairness because detection of misaligned pulsars in radio is, by definition, not possible, while the detection of their associated PWNe is not hindered, as the cosmic rays that generate the gamma-ray emission most likely follow an isotropic distribution. The exclusion of pulsars without known distance is to avoid having an outlier that drags the fit extremely far from the best-fit parameters because of a few sources whose upper limit is $2\textendash3$ orders of magnitude too high/low. These selection criteria leave us with $17$ pulsars. To compute the upper limit for each PWN, we use the sensitivity presented in \cite{lhaaso_catalog} multiplied by $4\pi d^2$ and take into account the loss of sensitivity for extended source. We also account for the fact that different regions of the sky have different exposures by multiplying the sensitivity by a declination-dependent attenuation factor \citep{sommers}. To estimate the best-fit parameters $\left(\alpha, \beta,\sigma\right)$ we perform Bayesian inference using a weakly informative (nearly flat) prior and then sample the posterior distribution.

To better illustrate how the censored linear regression works and why it is more reliable for our study, we show figure \ref{fig:fits} which presents the result of a fit on ATNF and KM2A data using a standard linear regression (magenta line) compared to the fit resulting from a censored linear regression (green line). Figure \ref{fig:fits} shows that a standard linear regression performed solely on the $17$ PWNe detected by KM2A leads to a fitted line that mostly replicates  and accounts for these $17$ sources and overpredicts the flux of tens to hundreds of undetected PWNe, to the point where the generated fluxes of undetected PWNe may be much higher than the detection threshold. This leads to a situation inconsistent with the current data. On the other hand, the censored linear regression is driven by the upper limits on the flux of unresolved PWNe for small spindown powers for which there is almost no detection. As a result, the fitted line is dragged down, below the detection threshold of most sources, which does not incur any conflict with the data. For higher spindown powers, where more detections are available, the fit is made in such a way to reproduce the number of sources detected by KM2A.

\section{Revisiting the associations of the LHAASO catalog} \label{sec:off_beamed_sec}
The main goal of this section is to estimate the best-fit values for the parameters $\left(\alpha,\beta,\sigma\right)$ of our fit of the intrinsic emission of sources above $10\,\rm{TeV}$ as a function of their spindown power. However, as previously mentioned, because of the very limited statistics, the result of the fits is very sensitive to the data, and particularly to any potential misidentification in \cite{lhaaso_catalog}. To handle this, we first check for any tension between the number of aligned pulsars reported in \cite{lhaaso_catalog} and the beaming fractions usually quoted in the literature. Then we present the procedure introduced by \cite{Mattox_1997} and adapt it to our study to reliably identify any misidentification. Finally, we use this procedure on the LHAASO catalog to identify and handle misidentified sources, and subsequently estimate the best-fit parameters of our model.

\subsection{A tension between the LHAASO catalog and the beaming fraction of pulsars}\label{subsec:problem}
There are $31$ sources detected by KM2A in the LHAASO catalog that have a pulsar association. Among these, $18$ are radio-loud, while the other $13$ are associated to radio quiet pulsars assumed to be misaligned. If we assume that all of the $18$ source associations with aligned pulsars made by LHAASO are correct and that the beaming fraction of pulsars is of the order of $20\%$, as suggested by the fits of \cite{off_beamed} on radio data, we obtain $\sim180$ detectable sources in the FOV of LHAASO. This rough estimate, which takes into account both aligned pulsars that are undetected and misaligned pulsars, is in tension with the number of $75$ sources reported by KM2A. 

In order to demonstrate this result quantitatively we run $1000$ simulations in which we generate randomly different lists of sources following the procedure presented in section \ref{sec:methods}. As previously mentioned, we assume that all of the $18$ associations of LHAASO are correct and therefore perform our fit of the total flux above $10\,\rm{TeV}$ as a function of the spindown using each one of the $18$ sources that has a measured distance. This result leads to the fitted expression
\begin{equation}\label{base_fit_prior}
    \log\left(\frac{4\pi d^2F\left(E>10\,\rm{TeV}\right)}{\rm{s}^{-1}}\right) = 0.990\log{\left(\frac{\dot{E}}{\rm{erg}\,\rm{s}^{-1}}\right)} -5.006
\end{equation}
to which we add a Gaussian noise with a standard deviation $\sigma=1.349$ to account for the scattering of the data around the fit. Once our source lists have been generated we test each source in the FOV of LHAASO to see whether it is detectable by KM2A or not.

Following the procedure described above, we find that with this setup we get on average $160.4\pm17.6$ sources detectable by KM2A, which represents a $\sim5\sigma$ discrepancy with the $75$ sources reported in \cite{lhaaso_catalog}. Moreover, the discrepancy becomes even larger if some of the $13$ radio-quiet pulsars to which LHAASO sources are associated are considered as aligned pulsars undetected in radio instead of misaligned pulsars. To reduce the discrepancy one could decrease the fraction of misaligned pulsars by renormalizing the formula of \cite{off_beamed} to a smaller percentage of the total population of pulsars in the Galaxy. However, going down to $70.2\pm11.3$ detectable PWNe requires a fraction of misaligned pulsars of $\sim60\%$, already too low or at the lower end of what is usually quoted in the literature \citep{60percentmin, 75percentmin}. With these constraints, the only solution is to consider that some of the sources detected by KM2A may not be associated to the correct pulsar. Such a situation is not surprising and, in fact, even expected at UHE where the large extension of sources and the limited pointing accuracy of water-Cherenkov detectors, like LHAASO or HAWC, make chance coincidences much more frequent \citep{75percentmin}.

\subsection{Probability of correct identification}\label{subsec:mattox}
In order to robustly report LHAASO sources misidentified to ATNF pulsars we propose to adapt the procedure introduced by \cite{Mattox_1997} to associate EGRET gamma-ray sources with extragalactic radio sources which gives the probability of correct identification for each source candidate. In the context of \cite{Mattox_1997} the probability that a radio source at an angular distance $r$ of the EGRET source is the correct association reads as:
\begin{equation}\label{mattox_test}
    p\left(\rm{id}|r\right) = \frac{\left[\eta/\left(1-\eta\right)\right]\rm{LR}}{\left[\eta/\left(1-\eta\right)\right]\rm{LR}+1}
\end{equation}
where $\eta\equiv p\left(\rm{id}\right)$ is the prior probability for a radio source to be the counterpart of the gamma-ray source and $\rm{LR}\equiv3\frac{r_0^2}{\Psi^2}\exp{\left(-r^2\left(3\Psi^{-2}-r_0^{-2}\right)\right)}$ is the likelihood ratio with $r_0$ being a characteristic angular distance between two different radio sources and $\Psi$ is, in the case of a circular error region, the $95\%$ confidence radius and $p\left(\rm{id}|r\right)$ is the posterior probability of the source to be the correct association knowing its angular distance from the EGRET source. This approach is based on the Bayes' theorem and works as follows. First, a prior probability to be the correct identification is computed based on data, a theoretical model or, more generally, any information available on the potential of a given source to be the correct radio counterpart of the EGRET source. After that, the prior probability is updated leveraging the Bayes' theorem to obtain the posterior probability. In addition to encoding the information carried by the prior probability, the posterior probability also accounts for contributions arising from the angular distance between the radio source and the EGRET source, which makes physical association less likely as it grows, the level of source crowding in the region and the pointing accuracy of the detector, which make confusion more likely to happen. This approach has the benefit to be comprehensive and, therefore, more reliable to estimate the probabilities of correct association.

In the context of their study, \cite{Mattox_1997} construct their prior probability $\eta$ as the probability of a radio source to be detected by EGRET as a function of its flux at $5\,\rm{GHz}$. For their characteristic angular distance, they define it as $r_0 \equiv\left[\pi\rho\left(S_5,\alpha\right)\right]^{-1/2}$ where $\rho\left(S_5,\alpha\right)$ is the number density of radio sources with a flux density higher than the flux density of the radio assessed as a counterpart and a spectrum at least as flat. Finally, $\Psi$ is defined as the $95\%$ confidence radius for the position of the EGRET source in the case of a circular error region.

In the case of Galactic pulsars some adjustments need to be made to the approach of \cite{Mattox_1997}. In our context, we set $r_0 \equiv\left[\pi\rho\left(\dot{E}/d^2\right)\right]^{-1/2}$ where $\rho\left(\dot{E}/d^2\right)$ is the local density of pulsars with $\dot{E}/d^2$ higher than that of the pulsar association candidate. This (simple) definition reflects the facts that a pulsar with a higher spindown power have a more substantial energy budget for particle acceleration and subsequent gamma-ray production, and that the emitted gamma-ray flux, whatever its value is, is diluted by a factor proportional to $d^2$. For $\Psi$ we take the $68\%$ containment region of the gamma-ray emission, which we consider to be the source extension. This choice is made to accommodate for the very large extension of sources at UHE (usually larger than the error region) and also to account for the fact that the centroid of the gamma-ray emission can be relatively far from the location of a source depending on the geometry of Galactic magnetic field at the location of the source \citep{offset}. Finally, we take as the prior probability $\eta$ the cumulative probability for a given pulsar with known spindown power and distance to create an UHE gamma-ray flux at least as high as the one observed. We define such a probability based on equation (\ref{base_fit_prior}). It reads as
\begin{equation}\label{cumu}
    p\left(\rm{id}\right) \equiv \int_{\tilde{F}}^{+\infty}{\mathcal{N}\left(x|\mu_0, \sigma^2_0\right)}\,dx = 0.5\times\operatorname{erfc}\left(\frac{\tilde{F} - \mu_0}{\sigma_0\sqrt{2}}\right)
\end{equation}
where $\mathcal{N}\left(x|\mu_0, \sigma^2_0\right)$ is a Gaussian of mean value $\mu_0$ and standard deviation $\sigma_0$, $\mu_0=0.990\log{\left(\frac{\dot{E}}{\rm{erg}\,\rm{s}^{-1}}\right)} - 5.006$ is the expected mean emission based on our model, $\sigma_0 = 1.349$ is the intrinsic scattering (standard deviation) predicted by our model, $\tilde{F} =\log\left(\frac{4\pi d^2F\left(E>10\,\rm{TeV}\right)}{\rm{s}^{-1}}\right)$ is the logarithm of the flux measured by KM2A above $10\,\rm{TeV}$ multiplied by $4\pi$ times the square of the distance of the pulsar candidate, $x$ is the integration variable, and $\operatorname{erfc}$ denotes the complementary error function. This form of the prior probability is physically well motivated. When the spindown power increases, the prior probability also increases which means that a higher energy budget makes the observed flux more likely to happen. On the other hand, when the distance increases the prior probability to have the observed flux decreases, as the intrinsic emission emitted by the source is more diluted. Finally, when the observed flux increases, the prior probability to observe it decreases, as it penalizes source with lower energy budget, which are less likely to be the one generating such a high flux. Moreover, this choice is qualitatively consistent with theoretical expectations. Indeed, it predicts a prior probability for the Crab to be the correct association to the LHAASO source 1LHAASO J0534+2200u at the level of $\sim75\%$ as expected for the Crab, which is very powerful and relatively close to the Sun. On the other hand, it predicts a prior probability of correct association at the level of $\sim1\%$ for the LHAASO source 1LHAASO J1914+1150u whose reported association is a pulsar located at more than $14\,\rm{kpc}$ with a moderate spindown power around $5\times10^{35}\,\rm{erg}\,\rm{s}^{-1}$. See table \ref{tab:assoc} for the numbers just presented.

\subsection{Resolving the tension between the LHAASO catalog and standard pulsar beaming fractions } \label{subsec:work}
In this subsection we apply the procedure introduced in subsection above to assess the robustness of the associations presented in the LHAASO catalog \citep{lhaaso_catalog}. We test several scenarios assuming different beaming fractions corresponding to percentages of misaligned pulsars ranging from $0\%$ to $\sim80\%$. In particular, we investigate configurations of the Galaxy in which there are $0\%$, $50\%$, $67\%$ and $80\%$ misaligned pulsars. For each case, we run $1000$ simulations where we generate different lists of sources following the procedures introduced in section \ref{sec:methods}. After that, we use the generated lists to compute the characteristic angular distance $r_0$ for each source list. For the prior probability $\eta$, the angular distance $r$ between the LHAASO source and pulsar candidate, and the $68\%$ containment region of the gamma-ray emission $\Psi$ which only depend on the source investigated are computed at the beginning and unchanged afterwards. Using these parameters, we then compute the posterior probability of correct identification $p\left(\rm{id}|r\right)$ for each source reported by KM2A and for each percentage of misaligned pulsars in the simulation. We present the results of this procedure in table \ref{tab:assoc}. In addition, we show in table \ref{tab:assoc} the spindown power and the distance of the pulsar candidate, we mention (if any) the corresponding source observed by Imaging Air Cherenkov Telescopes (IACTs) which have a much better angular resolution, and we mention whether the pulsar is aligned or misaligned. It is worth mentioning that at this point we have not yet investigated the gamma-ray emission of the sources or the ability of KM2A to detect them. We have only provided posterior probability of correct association.

\begin{deluxetable*}{lccccccccc}
\tablewidth{0pt}
\tablecaption{Summary of the properties of the pulsars associated to the LHAASO sources reported by KM2A, together with the probabilities of correct association for different percentages of misaligned pulsars in the simulation. \label{tab:assoc}}
\tablehead{
\colhead{} & \colhead{} & \colhead{} & \colhead{} & \colhead{} & \multicolumn{4}{c}{Fraction of misaligned pulsars}& \colhead{}\\
\colhead{} & \colhead{} & \colhead{} & \colhead{} & \colhead{} & \colhead{$0\%$} & \colhead{$50\%$} & \colhead{$67\%$} & \colhead{$80\%$}& \colhead{}\\
\cline{6-9}
\colhead{Name} & \colhead{IACT} & \colhead{$\dot{E}$} & \colhead{$d$} & \colhead{$p\left(\rm{id}\right)$} & \multicolumn{4}{c}{$p\left(\rm{id}|r\right)$}& \colhead{Aligned?}\\
\colhead{} & \colhead{} & \colhead{($10^{36}\,\rm{erg}\,\rm{s}^{-1}$)} & \colhead{($\rm{kpc}$)} & \colhead{($\%$)} & \colhead{($\%$)} & \colhead{($\%$)} & \colhead{($\%$)} & \colhead{($\%$)} & (Y/N)
}
\startdata
1LHAASO J0249+6022 & --- & $0.21$ & $2.0$ & $7.91$ & $93.74$ & $91.19$ & $86.61$ & $77.05$ & Y\\
1LHAASO J0359+5406$^{\ddagger}$ & --- & $1.3$ & $3.45$ & $12.30$ & $97.84$ & $97.51$ & $96.59$ & $93.59$& Y\\
1LHAASO J0534+2200u & Crab Nebula & $450$ & $2.0$ & $74.73$ & $99.99^{\ast}$ & $99.99^{\ast}$ & $99.99^{\ast}$ & $99.99^{\ast}$ & Y\\
1LHAASO J0542+2311u & --- & $0.041$ & $1.56$ & $1.75$ & $25.55$ & $24.49$ & $22.81$ & $19.81$& Y\\
1LHAASO J0631+1040 & --- & $0.17$ & $2.1$ & $12.67$ & $98.28^{\ast}$ & $98.28^{\ast}$ & $98.28^{\ast}$ & $98.28^{\ast}$ & Y\\
1LHAASO J1740+0948u & --- & $0.23$ & $1.23$ & $29.52$ & $91.91$ & $91.89$ & $91.74$ & $91.71$& Y\\
1LHAASO J1809-1918u & --- & $1.8$ & $3.27$ & $4.78$ & $87.66$ & $87.06$ & $85.04$ & $81.44$& Y\\
1LHAASO J1825-1337u & HESS J1825-137 & $2.8$ & $3.61$ & $6.40$ & $92.15^{\ast}$ & $91.84^{\ast}$ & $90.31^{\ast}$ & $87.63^{\ast}$ & Y\\
1LHAASO J1908+0615u & --- & $2.8$ & $2.37$ & $16.83$ & $96.74^{\ast}$ & $96.49^{\ast}$ & $95.32^{\ast}$ & $93.11$& Y\\
1LHAASO J1912+1014u & --- & $2.9$ & $4.61$ & $14.65$ & $83.70$ & $82.63$ & $78.78$ & $72.91$& Y\\
1LHAASO J1914+1150u & --- & $0.54$ & $14.01$ & $1.58$ & $26.82$ & $22.66$ & $18.16$ & $12.45$& Y\\
1LHAASO J1928+1746u & --- & $1.6$ & $4.34$ & $18.35$ & $98.28^{\ast}$ & $98.08^{\ast}$ & $97.49^{\ast}$ & $96.33^{\ast}$& Y\\
1LHAASO J1929+1846u & HESS J1930+188 & $12$ & $7.0$ & $29.68$ & $92.65^{\ast}$ & $92.13^{\ast}$ & $89.23^{\ast}$ & $84.70^{\ast}$ & Y\\
1LHAASO J2005+3415 & --- & $0.58$ & $10.78$ & $2.29$ & $23.55$ & $18.81$ & $13.89$ & $8.86$& Y\\
1LHAASO J2005+3050 & --- & $0.22$ & $6.04$ & $3.58$ & $45.03$ & $37.78$ & $29.11$ & $19.21$& Y\\
1LHAASO J2020+3649u & --- & $3.4$ & $1.80$ & $29.25$ & $99.83^{\ast}$ & $99.83^{\ast}$ & $99.79^{\ast}$ & $99.70^{\ast}$ & Y\\
1LHAASO J2031+4127u & VER J2032+414  & $0.15$ & $1.33$ & $7.56$ & $93.02^{\ast}$ & $92.14^{\ast}$ & $89.75^{\ast}$ & $84.99^{\ast}$ & Y\\
1LHAASO J2228+6100u & --- & $22$ & $3.0$ & $35.07$ & $98.64^{\ast}$ & $98.59^{\ast}$ & $98.34^{\ast}$ & $97.89^{\ast}$ & Y\\
1LHAASO J2238+5900 & --- & $0.89$ & $2.83$ & $9.37$ & $92.39$ & $89.28$ & $83.51$ & $72.86$& N\\
1LHAASO J0007+7303u & VER J0006+729 & $0.45$ & $1.40$ & $11.68$ & $99.42$ & $99.42$ & $99.39$ & $99.29$& N\\
1LHAASO J0622+3754$^{\dagger}$ & --- & $0.027$ & --- & $13.75$ & $96.81$ & $96.81$ & $96.73$ & $96.52$& N\\
1LHAASO J0635+0619 & --- & $0.12$ & $1.35$ & $10.03$ & $88.26$ & $88.26$ & $88.26$ & $88.26$& N\\
1LHAASO J0634+1741u & Geminga & $0.032$ & $0.19$ & $17.26$ & $91.61$ & $91.61$ & $91.61$ & $91.61$& N\\
1LHAASO J1813-1245 & --- & $6.2$ & $2.63$ & $28.07$ & $99.22$ & $99.18$ & $99.05$ & $98.83$& N\\
1LHAASO J1825-1256u & --- & $3.6$ & $1.55$ & $24.34$ & $99.45$ & $99.43$ & $99.33$ & $99.15$& N\\
1LHAASO J1837-0654u & HESS J1837-069 & $5.5$ & $6.60$ & $7.02$ & $85.24$ & $83.94$ & $79.89$ & $73.14$& N\\
1LHAASO J1839-0548u$^{\dagger}$ & --- & $5.7$ & --- & $52.44$ & $99.69$ & $99.69$ & $99.69$ & $99.69$& N\\
1LHAASO J1848-0001u$^{\dagger}$ & HESS J1849-000  & $9.8$ & --- & $51.27$ & $99.97$ & $99.96$ & $99.96$ & $99.96$& N\\
1LHAASO J1954+2836u & --- & $1.0$ & $1.96$ & $38.12$ & $99.80$ & $99.79$ & $99.73$ & $99.61$& N\\
1LHAASO J1959+2846u & --- & $0.34$ & $1.95$ & $19.52$ & $95.46$ & $95.03$ & $93.30$ & $89.82$& N\\
1LHAASO J2028+3352$^{\dagger}$ & --- & $0.035$ & --- & $47.79$ & $92.63$ & $92.58$ & $92.49$ & $92.03$& N\\
\enddata
\tablecomments{Column 1 lists the names of the PWNe detected by KM2A. Column 2 lists the counterpart of the detected sources seen by an Imaging Air Cherenkov Telescope (IACT) that ensures that the association is correct (except for Geminga which is confidently associated). If no counterpart is available, then nothing is listed. Columns three and four give respectively the spindown power and the distance (when available) of the pulsar to which the source is associated. Column five gives the prior probability of correct association based on equation \ref{base_fit_prior}. Columns from six to nine give the posterior probability for $0\%$, $50\%$, $67\%$ and $80\%$ misaligned pulsars. Column ten states whether the pulsar is aligned (Y) or misaligned (N). For the radio detection, we follow the ATNF catalog, except for Geminga that we treat as radio-quiet due to the current debate about around its radio detection. Probabilities with the $^{\ast}$ means that the pulsar has been selected for the subsequent fit. Names with the $^{\dagger}$ symbol means that the pulsar does not have a known distance and, therefore, the minimal distance defined by the ratio of the physical and angular extensions is assumed. The probabilities for these sources are reported for the sake of completeness but are not reliable. Such sources have been excluded from all fits. PSR J0359+5414 (associated to 1LHAASO J0359+5406) is marked with $\ddagger$ because it has been detected in radio recently \citep{ding_2025}, after we have completed the fits done for this paper.}
\end{deluxetable*}

In order to conclude on the correctness of associations we need to define a threshold probability above which we accept the association and under which we reject it. Setting such a threshold is somewhat arbitrary and depends on the context and needs of the study. Here we choose a threshold probability at $95\%$. In our context, a higher threshold may become too stringent and exclude an unreasonable number of sources, to the point where even censored regression models can no longer cope with the lack of statistics. A lower threshold would, on the other hand, allow too many misidentified sources to be used in the fits, which may mislead the estimation of the best-fit parameters and cause the loss of self-consistency in the results presented below. Besides the chosen threshold, we also take into account any concurrent observation of the LHAASO sources by IACTs such as HESS or VERITAS. More specifically, if a LHAASO source is also observed and associated to a pulsar by an IACT, which has a much better angular resolution, we consider the source to be the correct association, regardless of its posterior probability. Following these criteria, we have reported the sources that we consider correctly associated for each percentage of misaligned pulsars in table \ref{tab:assoc} by marking their posterior probability with an asterisk.

\begin{figure}
    \centering
    \includegraphics[width=\linewidth]{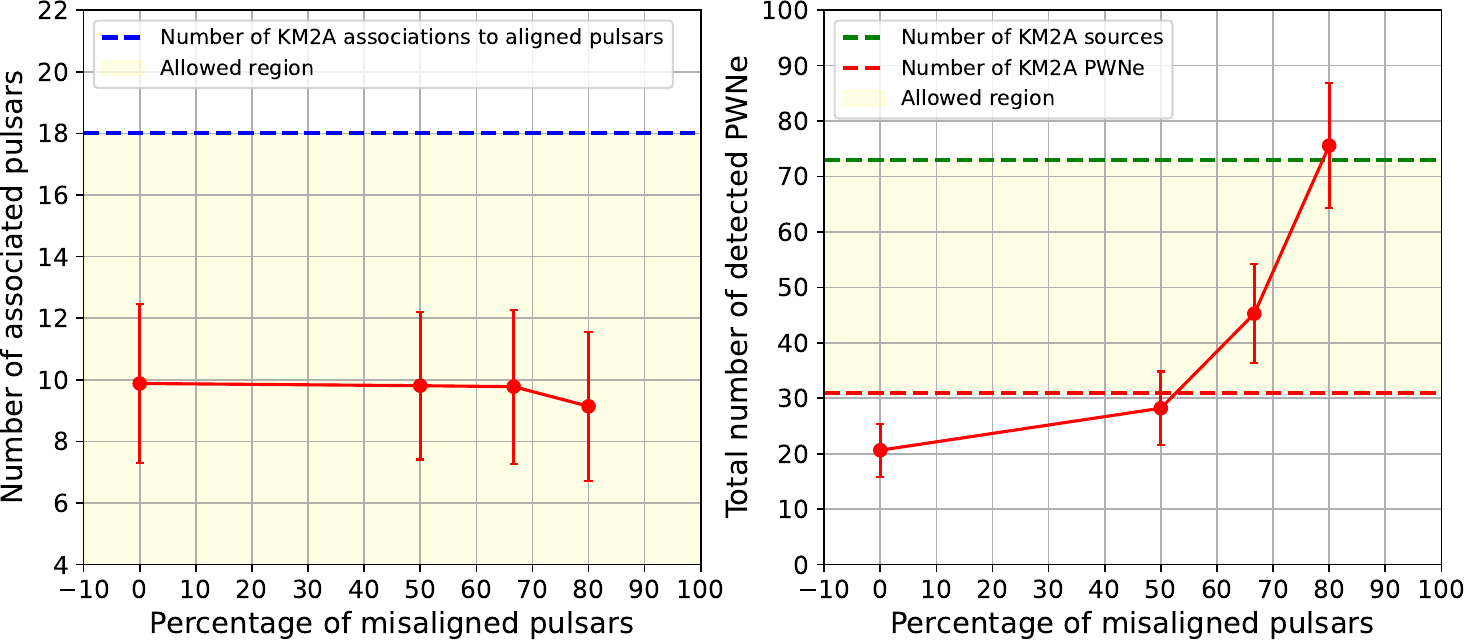}
    \caption{The left panel represents the number of detectable PWNe whose parent pulsar is part of the ATNF catalog as a function of the percentage of misaligned pulsars in the simulation. The right panel represents the total number of detected PWNe (whose parent pulsar is either in the ATNF catalog or synthetic) as a function of the percentage of misaligned pulsars in the simulation. In both panels the data points are the average over $1000$ simulations and the error bars represent $1\sigma$ (one standard deviation) over $1000$ simulations.}
    \label{fig:tensions}
\end{figure}

After that, we repeat the fit that allowed to estimate the best-fit parameters in equation (\ref{base_fit_prior}), but this time excluding the sources that have been associated to the wrong pulsar based on our selection criteria. Finally, using the best-fit parameters estimated from the updated fit, we repeat the procedure described in section \ref{sec:methods} and run $1000$ simulations. For each simulation we then compute the total number of detectable PWNe and the number of detectable PWNe whose parent pulsar is part of the ATNF catalog. To conclude on the detectability of a source, here again, we use the sensitivity presented in \cite{lhaaso_catalog} with the adequate corrections. The results of this procedure are presented in figure \ref{fig:tensions}. 

In figure \ref{fig:tensions} the left panel represents the number of PWNe detectable by KM2A and whose parent pulsars are part of the ATNF catalog as a function of the fraction of misaligned pulsars in the simulation. The blue line shows the maximum number of $18$ PWNe. The right panel represents the total number of PWNe detectable by KM2A as a function of the fraction of misaligned pulsars in the simulation. The red (green) line shows the minimum (maximum) number of detections consistent with KM2A observations.  In both panels the yellow shaded area shows the allowed region of the parameters space, the data points represent the average number of detections over $1000$ simulations and the error bars represent $1\sigma$ (one standard deviation) over $1000$ simulations. Comparing the number of detected PWNe whose parent pulsar is part of the ATNF catalog in the left panel of figure \ref{fig:tensions} and the number of LHAASO sources correctly associated shows that our modeling is self-consistent. For all percentages of misaligned pulsars, the number of PWNe that are detected in our simulations and whose parent pulsar is part of the ATNF catalog is consistent with the number of sources that we found to be correctly associated (see table \ref{tab:assoc}) and that we used for the fit. On the other hand, the right panel of figure \ref{fig:tensions} shows that fractions of misaligned pulsars that are smaller than $\sim50\textendash60\%$ are not compatible with the KM2A observations, as the number of detectable PWNe is underpredicted. On the other hand, we see that percentages higher than $\sim80\%$ overpredict the number of detectable PWNe by KM2A which sets an upper bound on the fraction of misaligned pulsars in the Galaxy. Reasonable values for the fraction of misaligned pulsars lay in the range $\sim60\textendash80\%$ which is compatible with the information present in table \ref{tab:assoc} which shows $9$ over $31$ sources that are associated to an aligned pulsar. For the rest of this work we consider that the fraction of misaligned pulsars is $\sim80\%$ following \cite{off_beamed} which is consistent with the LHAASO catalog, as it is within the statistical fluctuations but requires that most of the unidentified sources reported in the LHAASO catalog are PWNe associated to a misaligned pulsar that has not yet been discovered (see the green dashed line in the right panel of figure \ref{fig:tensions}). In this case the fit of the total flux above $10\,\rm{TeV}$ of equation (\ref{base_fit_prior}) becomes:
\begin{equation}\label{best_fit}
    \log\left(\frac{4\pi d^2F\left(E>10\,\rm{TeV}\right)}{\rm{s}^{-1}}\right) = 0.942\log{\left(\frac{\dot{E}}{\rm{erg}\,\rm{s}^{-1}}\right)} -3.894
\end{equation}
to which we add a Gaussian noise with a standard deviation $\sigma=1.197$.

\section{Application to the diffuse gamma-ray background}\label{sec:background}
In this section we propose to constrain the contribution of unresolved PWNe to the UHE Galactic diffuse gamma-ray background using our model. The UHE Galactic diffuse gamma-ray background is the UHE gamma-ray emission of our Galaxy arising from the interactions of the diffuse sea of very-high-energy cosmic rays with the different components of the interstellar medium. The study of the diffuse background is very important to understand the propagation mechanisms of cosmic rays as well as the nature and abundance of Galactic PeVatrons \citep{me1, me3}. In the energy range considered in this work ($\gtrsim1\,\rm{TeV}$) the contribution of leptonic cosmic rays to the so-called true diffuse background is expected to be negligible because of the very short cooling time of leptons. However, a poorly constrained level of leptonic contamination, notably from PWNe, is still possible in the form of photons from unresolved PWNe that would artificially increase the level of the diffuse background. To constrain the contribution of unresolved PWNe to the Galactic diffuse gamma-ray emission measured by LHAASO \citep{lhaaso_diffuse_2} above $1\,\rm{TeV}$ we mainly follow, with some improvements, the procedure presented in \cite{me2} that we summarize here. 

We simulate $1000$ configurations of the Galaxy. In each realization, we generate a new list of pulsars based on the modeling presented in section \ref{sec:methods} assuming a fraction of misaligned pulsars $\epsilon\sim80\%$. For the gamma-ray emission of the associated PWNe we also follow section \ref{sec:methods} with the fitted parameters presented in equation (\ref{best_fit}) for the energy range $25\textendash1000\,\rm{TeV}$, while for the energy range $1\textendash25\,\rm{TeV}$ we repeat the fits done for KM2A using the parameters of WCDA and obtain:
\begin{equation}\label{best_fit_wcda}
    \log\left(\frac{4\pi d^2F\left(E>1\,\rm{TeV}\right)}{\rm{s}^{-1}}\right) = 1.029\log{\left(\frac{\dot{E}}{\rm{erg}\,\rm{s}^{-1}}\right)} -5.682
\end{equation}
to which we add a Gaussian noise with a standard deviation $\sigma=1.152$.

\begin{figure}[h!]
    \centering
    \includegraphics[width=\linewidth]{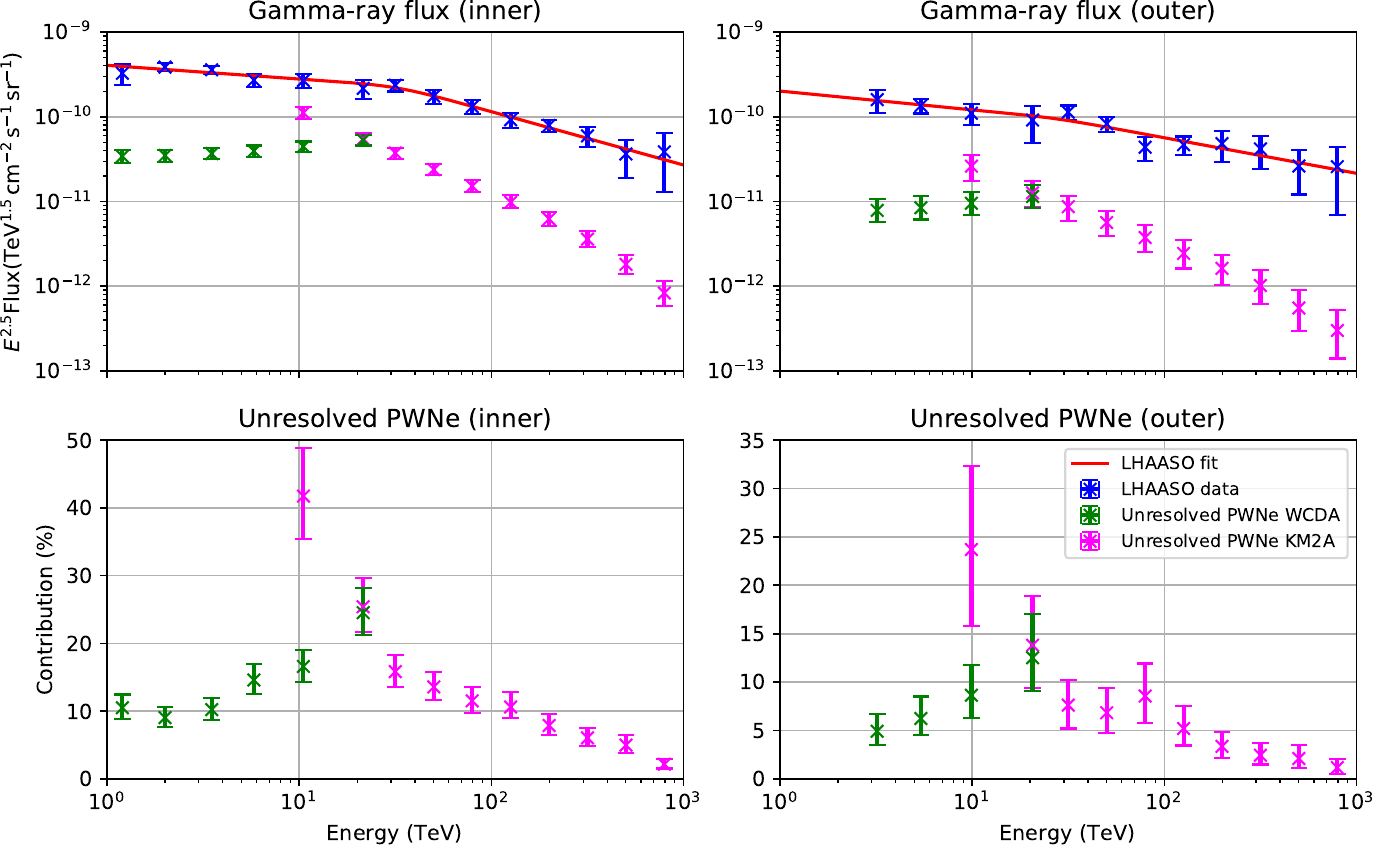}
    \caption{Contribution of unresolved PWNe to the Galactic diffuse gamma-ray background. The upper left (right) panel shows the diffuse gamma-ray flux of \cite{lhaaso_diffuse_2} and our prediction for the gamma-ray flux from PWNe for the inner (outer) Galaxy using the same masks as LHAASO. The lower left (right) panel shows the contribution (in $\%$) of unresolved PWNe to the diffuse gamma-ray flux of the inner (outer) Galaxy. The data points represent the average over $1000$ simulations and the error bars represent $1\sigma$ (one standard deviation) over $1000$ simulations.}
    \label{fig:diffuse}
\end{figure}

In our simulation, we ensure that all sources remain undetectable and then use the same masks as \cite{lhaaso_diffuse_2} to compute the diffuse gamma-ray flux. We take into account the absorption of gamma-rays during their propagation, due to collisions with cosmic microwave background photons, dust emission and starlight by using the absorption probabilities pre-computed by \cite{lipari_att}. We present our results in figure \ref{fig:diffuse}.

Figure \ref{fig:diffuse} shows the contribution of unresolved PWNe to the Galactic diffuse gamma-ray background. The upper left (right) panel shows the diffuse gamma-ray flux of \cite{lhaaso_diffuse_2} and our prediction for the gamma-ray flux from PWNe for the inner (outer) Galaxy using the same masks as LHAASO. The lower left (right) panel shows the contribution (in $\%$) of unresolved PWNe to the diffuse gamma-ray flux of the inner (outer) Galaxy. In all panels our predictions for WCDA are presented in green and our predictions for KM2A are presented in magenta. The data points represent the average over $1000$ simulations and the error bars represent $1\sigma$ (one standard deviation) over $1000$ simulations.

From figure \ref{fig:diffuse} we see that the contribution of PWNe to the Galactic diffuse gamma-ray emission is, in general, small for both the inner and outer Galaxy over the entire energy range $1\textendash1000\,\rm{TeV}$. For the inner Galaxy it does not exceed $\sim30\%$ around $25\,\rm{TeV}$ and is almost always less than $20\%$, while for the outer Galaxy it almost always contributes for less than $10\%$ with a peak around $20\%$, also around $25\,\rm{TeV}$. This result comes as a refinement of our previous findings \citep{me2} that was aiming at giving an estimate of the upper limit for this contribution. The difference between our two results is due to the introduction of a maximum energy for sources through the formula of \cite{e_max}, and to our improved fitting procedure, leveraging a censored analysis, to derive the relationship between the spindown power and the gamma-ray flux of sources.
\begin{figure}[h!]
    \centering
    \includegraphics[width=\linewidth]{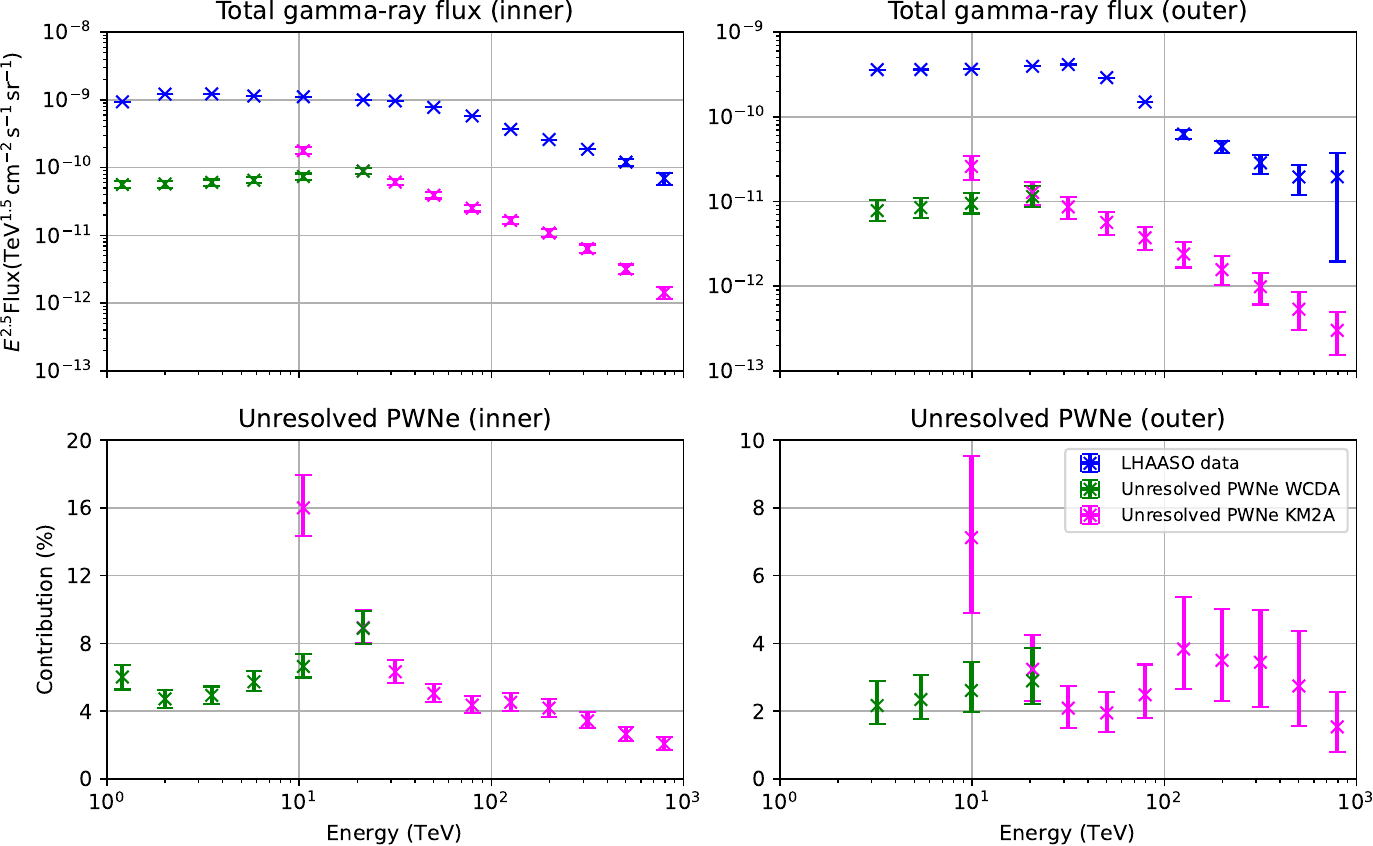}
    \caption{Contribution of unresolved PWNe to the total Galactic gamma-ray flux. The upper left (right) panel shows the total gamma-ray flux of \cite{lhaaso_diffuse_2} and our prediction for the total gamma-ray flux from PWNe for the inner (outer) Galaxy. The lower left (right) panel shows the contribution (in $\%$) of unresolved PWNe to the total gamma-ray flux of the inner (outer) Galaxy. The data points represent the average over $1000$ simulations and the error bars represent $1\sigma$ (one standard deviation) over $1000$ simulations.}
    \label{fig:total}
\end{figure}

For the sake of completeness, we also present in figure \ref{fig:total} the contribution of unresolved PWNe to the total flux measured by LHAASO, including diffuse and source components, without any masks. From figure \ref{fig:total} we see that the flux of unresolved PWNe without the masks is higher than the flux with the masks by a factor of $\sim1.7$ in the inner Galaxy and roughly the same in the outer Galaxy. This is expected for two reasons. First, most sources are located in the inner Galaxy, while very few of them are located in the outer Galaxy. This means that for a given angular area masked in the inner and the outer Galaxy, one should expect more sources to be masked and have their contribution removed in the inner Galaxy. Second, the fact that there are very few sources in the outer Galaxy implies that there are also very few detectable sources in the outer Galaxy. As a result, the region masked by LHAASO is not very large, leading to a minimal impact of the masks in that region. Finally, concerning the contribution of unresolved PWNe to the diffuse flux of LHAASO without masks, it is difficult to reach a definitive conclusion. As shown in figure \ref{fig:total} the contribution of PWNe to the total flux without masks is very small, usually smaller than $10\%$. Therefore, the contribution of unresolved PWNe to the diffuse background depends on the still unknown relative level of the diffuse background itself compared to the level of the flux of resolved sources. Indeed, if the total flux of LHAASO is largely dominated by the resolved source component, then we may expect that without the masks, the contribution of unresolved PWNe to the diffuse background will be important, alternatively, it will be completely negligible.

\section{Discussion and Conclusion}\label{sec:conclusion}
In this work, we have proposed a new data-driven parametrization of the UHE gamma-ray emission of PWNe. To build our model, we have used the information available in the ATNF and the LHAASO catalogs, to extract the spatial and age distributions of pulsars and the gamma-ray emission from PWNe. After that, we have used this information to construct our model of the UHE gamma-ray emission of PWNe in a self-consistent way to reproduce the data of the first LHAASO catalog and make a prediction of the contribution of unresolved PWNe to the UHE Galactic diffuse gamma-ray emission measured by LHAASO in the energy range of $1\textendash1000\,\rm{TeV}$.

While estimating the best-fit parameters of our model, we pointed out an apparent tension between radio and gamma-ray data in terms of predicted number of detectable PWNe. This tension can naturally be alleviated by identifying, through our adaptation of the procedure of \cite{Mattox_1997}, the LHAASO sources that have been mis-associated to aligned ATNF pulsars. Moreover with our model we have established that a substantial fraction of the unidentified sources in the LHAASO catalog need to be PWNe associated to yet undiscovered pulsars. Finally, using the best-fit parameters for our model, we have estimated the contribution of unresolved PWNe to the UHE Galactic diffuse gamma-ray emission, that we found to be in general minor to negligible.

One of the main challenges faced in this study is the limited data available in gamma-rays. This is a natural limitation arising from the difficulty to simultaneously satisfy all the physical conditions to achieve substantial particle acceleration to multi-$\rm{PeV}$ energies which results in very few detectable sources. To cope with this, we have introduced censored regression models that also include the information arising from the upper limits on unresolved sources. Such an approach led to physically coherent results and the prediction of an extremely small gamma-ray flux, far from the detection threshold, for most PWNe with a small spindown power ($\dot{E}\lesssim10^{34}\,\rm{erg}\,\rm{s}^{-1}$). Such a result can find further support in the future if longer observation times fail to reveal a substantial number of new PWNe associated to pulsars with $\dot{E}\simeq10^{34}\,\rm{erg}\,\rm{s}^{-1}$. Moreover, longer observation times will also help to reveal more sources and decrease the upper limits on unresolved sources, allowing to further refine our fits.

To reconcile the number of PWNe detected by LHAASO and associated to aligned pulsars with usual beaming fractions of pulsars, we have put forward the possibility that some of the associations made by LHAASO may be incorrect. We argue that this possibility is the only reasonable way to reconcile the LHAASO data with plausible fractions of misaligned pulsars around $\sim60\textendash80\%$.

In order to detect incorrect associations, we have adapted the procedure of \cite{Mattox_1997} to Galactic pulsars. Our adapted procedure relies on two arbitrary assumptions that are the form of the prior probability of identification and the acceptance threshold for the posterior probability of identification. We have constructed the prior probability based on the $18$ PWNe reported by LHAASO and presumably associated to aligned pulsars, while the correctness of these associations is not guaranteed. However, we argue that the constructed probability is self-consistent with our current physical understanding of the processes in play, providing results in agreement with physical expectations. Concerning the arbitrary threshold that we have assumed to be at $95\%$ we argue that this value is the best compromise that can be achieved. Indeed, in addition to ensure maximum purity of the statistical sample, our choice has been driven by the necessity to build a self-consistent model. Increasing the threshold would result in the exclusion of too many sources with high chance of correct associations to the point where even censored regression models will fail to provide reliable fits. On the other hand, decreasing our threshold would lead to an important contamination of our sample, resulting into fits that may possibly not reproduce the number of detections reported by LHAASO. Finally, the main weakness that we could not mitigate in our approach is the treatment of sources whose pulsar candidate does not have a measured distance. In this case, we predict a very high probability of correct association that is most likely not representative of the real situation. We stress that the probability of correct association for such sources are not reliable and that we have excluded them from all our fits.

During the final stages of preparation of this paper we found that a periodic radio signal has been detected from the previously radio-quiet pulsar J0359+5414 associated to 1LHAASO J0359+5406. For the sake of consistency we have calculated the prior probability defined by equation (\ref{base_fit_prior}) by including the corresponding LHAASO source. Although the prior probability has been slightly affected, it did not induce any appreciable change in the posterior probabilities and did not change the number of identifications that we consider to be likely to be correct. However, concerning the source 1LHAASO J0359+5406 itself it seems to reach the $95\%$ threshold of correct identification assuming $0\%$, $50\%$, and $67\%$ of misaligned pulsars in the simulation, but not for the case with $80\%$ misaligned pulsars. In the light of this new result, the fit for the former percentages of misaligned pulsars should be revised to include this source, while the fit assuming $80\%$ of misaligned pulsars, which is used to compute the contribution of unresolved PWNe to the diffuse background measured by LHAASO, remains unchanged.

If in the future more of the radio-quiet pulsars are eventually observed in radio, or some of disputable associations to LHAASO sources are confirmed, we expect the tension between radio and gamma-ray data to become too important in the case of $\sim80\%$ of misaligned pulsars, suggesting that a more reasonable value for the percentage of misaligned pulsars in the Galaxy would be around $\sim67\textendash75\%$. To take into account the impact of the inclusion of more sources, we have revisited the fit used in section \ref{sec:background} for the KM2A detector using all aligned pulsars with a probability of correct association higher than $80\%$ in the ninth column of table \ref{tab:assoc}. This leads to a total of $12$ \lq\lq correct\rq\rq associations used for the fit, including the LHAASO source 1LHAASO J0359+5406. In this case, we found an increase in the contribution of unresolved PWNe to the diffuse gamma-ray background at a level $\lesssim1.5$, still too small to be significant. However, such a fit would inevitably lead to overshooting the number of detections of KM2A and would require, in practice, to decrease the fraction of misaligned pulsars and the contribution of unresolved PWNe to the diffuse gamma-ray background.

In conclusion, we have presented a new data-driven procedure which provides a realistic parametrization of the UHE gamma-ray emission of PWNe with a high predictive power. We have shown that our parametrization enables to study various topics, such as the detectability of PWNe and their contribution ot the diffuse emission of our Galaxy, and expect that this modeling will help to better understand the nature of their emission as well as the mechanisms behind it.

\begin{acknowledgments}
The authors acknowledge the use of the ATNF catalog. Samy Kaci acknowledges funding from the Chinese Scholarship Council (CSC) and thanks Paolo Lipari for providing the absorption probabilities for gamma-rays. This work is supported by the National Natural Science Foundation of China under Grants Nos. 12350610239, and 12393853. This work was supported by the National Center for High-Level Talent Training in Mathematics, Physics, Chemistry, and Biology. The Work  of Dmitri Semikoz has received in part support under the program "Investissement d'Avenir" launched by the French Government and implemented by ANR, with the reference " ANR‐18‐IdEx‐0001 " as part of its program " Emergence ".
\end{acknowledgments}


\bibliography{sample701}{}
\bibliographystyle{aasjournalv7}



\end{document}